\documentclass[aps,prl,twocolumn,showpacs]{revtex4}

\usepackage{graphicx}
\usepackage{subfigure}

\begin{document}

\title{Vortex Chain States in a Ferromagnet/Superconductor Bilayer}

\author{Serkan Erdin\\
Department of Physics, Northern Illinois University, DeKalb, IL, 60115\\
\& Advanced Photon Source, Argonne National Laboratory,\\
 9700 South Cass Avenue,  Argonne, IL, 60439}


\begin{abstract}
The discrete vortex lattices in a 
ferromagnet/superconductor bilayer are studied when the ferromagnet has  
periodic stripe domains with an out-of-plane magnetization. The vortices are assumed to be 
situated periodically on 
chains in the  stripe domains. Only up to two vortex chains  per 
domain configurations are considered. When the domain period is fixed, the
threshold magnetization is calculated at which  
the transition from the Meissner state to the mixed state occurs. 
When the domain period is not fixed, the equilibrium domain size and 
vortex positions are calculated, 
depending on the domain's magnetization and the domain wall energy. 
In equilibrium, the vortices
in the neighbor domains are half-way shifted, while they are next to 
each other in the same domain.
\end{abstract}

\maketitle


\noindent PACS Number(s): 74.25.Dw, 74.25.Ha, 74.25.Qt, 74.78.-w



\section{introduction}

In recent years, heterostructures made of type II superconductors and 
ferromagnetic 
pieces have been the focus of studies both experimental and theoretical
\cite{pok-rev,serk-rev}. In these structures, the magnet  and the 
superconductor 
are 
separated by an oxide layer to avoid the proximity effect and spin 
difffusion that might lead to the supression of ferromagnetic(FM) and 
superconducting(SC) order 
parameters. Therefore, the interaction between the FM textures and the 
vortex matter in SC pieces is maintained only by the magnetic fields 
generated by the SC vortices and the magnet. This 
strong interaction 
not only gives rich physical effects that are not observed in 
individual subsystems, but also offers new devices that can be tuned by 
weak magnetic fields. 

One of the realizations of 
such 
heterostructures is a ferromagnetic/superconducting bilayer (FSB). 
In recent years, FSBs have drawn a great deal of attention both experimentally 
and theoretically. 
On the experimental side, nonsymmetric current-voltage characteristics
on La$_{0.66}$Sr$_{0.33}$MnO$_3$/YBa$_2$Cu$_3$O$_{7-\delta}$ were studied \cite{applphyslett}.
It has also been reported that stray fields from domain structures of an
FM film lead to a significant decrease 
of the SC critical temperature $T_c$ in a zero external field, 
whereas $T_c$ is enhanced under an  applied field \cite{lange}. 
The influence of magnetic domain walls on SC 
critical temperature was theoretically studied on the basis of the Ginzburg-Landau 
approach \cite{dwsuper}. On the theoretical side, much attention has been paid 
to the interplay between the vortex matter and the FM layer. Sonin  
studied the conditions for 
the penetration of a vortex near a magnetic domain wall \cite{sonin1}. 
Helseth {\it et al.} investigated the pinning of vortices near the 
magnetic domain walls \cite{helseth}. Recently, Laiho {\it et al.} 
investigated the vortex structures in the FSB, when the FM layer has 
domain structure with out-of-plane magnetization \cite{laiho}. Two 
possible vortex structures were shown to occur: vortices with alternating directions 
corresponding to the direction of the magnetization in FM domain and 
vortex semiloops connecting the adjacent FM domains.

Earlier Lyuksyutov and Pokrovsky \cite{pok2,pok7} noticed that, in a
bilayer consisting of homogeneous SC and
FM films with the magnetization normal to the plane, SC vortices occur
spontaneously in the ground state, even though the magnet does not
generate a magnetic field in the SC film.

In previous work, we presented a
theory of such vortex-generation instability and the resulting vortex
structures \cite{stripe}. We showed that, due to this instability, domains 
with
alternating magnetization and vortex directions occur in a FSB.  
 In that study, the domain structures were treated  in the continuum regime in which the 
domain
size was much larger than the effective penetration depth, $\Lambda = 
\lambda^2/d_{sc}$, where the London penetration depth $\lambda$ is much 
larger than the thickness of a SC film $d_{sc}$ 
\cite{abrikosov}. Under the continuum aproximation, the energy of 
stripe phase was found to be minimal. The equilibrium domain size
and the equilibrium
energy for the stripe structure were found as \cite{stripe}
                                                                                                                                                             
\begin{equation}
L_{eq}^{(str)}=\frac{\Lambda }{4}\exp
\left( \frac{\varepsilon _{dw}}
{4\tilde{m}^{2}}-C+1\right),
\label{L-eq-stripe}
\end{equation}
\begin{equation}
{U_{eq}^{(str)}}
=-\frac{16\widetilde{m}^{2}
{\cal A}}{\Lambda }\exp
\left( -\frac{\varepsilon _{dw}}{4\widetilde{m}^{2}}
+C-1\right),
\label{U-eq-stripe}
\end{equation}
\noindent where $\tilde m = m -\varepsilon_v/\phi_0$,  $\varepsilon_v =
(\phi_0^2/16 \pi^2 \Lambda) \ln (\Lambda/\xi)$ is the self-energy of a vortex,
$\varepsilon_{dw}$ is the domain wall energy per domain wall length, $\cal A$
is the domain's area and $C
\sim 0.577$ is the
Euler-Mascheroni constant. If $\varepsilon_{dw}\leq 4\tilde{m}^2$, the continuum approximation
becomes invalid, since $L_{eq}$ becomes on the order of or less than
$\Lambda$ (see Eq.(\ref{L-eq-stripe})). However, it can be
recovered
by considering the discrete lattice of vortices instead.

In this paper, the discrete lattice of vortices in the stripe domains is studied through  a 
method that works in both continuum and the 
discrete regimes. The method we use here is based on London-Maxwell equations
 and was developed elsewhere \cite{serkan1}.  Earlier, its extention  
to periodic systems for 
the case of 
square magnetic
dot arrays on a SC film was used \cite{ser-physica}. In this work, it is adapted 
to the 
discrete 
vortex 
lattices in
SC/FM bilayers. In doing so, we assumed that vortices and antivortices sit 
periodically on chains in the alternating domains of magnetization and 
vorticity. Recently, Karapetrov {\it et al.} observed vortex chains in  mesoscopic superconductor-normal metal 
hetrostructures by means of scanning tunneling spectroscopy techniques \cite{karap}. Vortex chains are also 
observed in anisotropic high $T_c$ superconductors (see review \cite{chains}). 
We first studied the vortex lattices in 
the FSB 
when the FM domain size is fixed. In this case, we investigated when the 
vortices appear spontaneously in the FSB, depending on the domain size and 
the magnetization strength of the domains. We found that one chain per domain configuration
appears first, followed by two chains per domain state
 upon  further increase of magnetization at the fixed domain
period. 
Next, we considered the case in which the domain size is not fixed. For this 
case, the following problems were studied; 
i) when the domain structures appear 
spontaneously;
ii) how the vortices and the antivortices are positioned on the
chains; iii) how the equilibrium domain size changes, depending on the
magnetization and the magnetic domain wall energy in the presence of the
vortices. In order to solve these 
problems, we
first proposed five different
configurations of the vortex and the antivortex chains, in which at most
two chains per stripe are considered.  Next, we calculated  their 
equilibrium
energies by means of numerical methods and found the most favorable case 
among them by minimizing their energies with respect to domain size and 
vortex positions.  
 Our calculations showed that,
in equilibrium structure, vortex
chains are half
shifted in the adjacent domain, while they are next to
each
other in the same  domains. The comparison of equilibrium energies of 
cases with one and
two chains per domains shows that, at lower values of  magnetization and
domain wall energy, the case with two chains is energetically favorable.
Additionally, the single-chain case does not win over the ones with two  
chains per domain in the equilibrium domain structures. 

The outline of this paper is as follows:
in the following section, the method for the discrete case and its 
application to configurations with single-vortex chain and double-vortex chains per 
domain is introduced. The third section is devoted to our results for vortex chain states in the domains with 
a fixed period. In the fourth section, we present 
our results on the  
proposed five  configurations of single-chain and double-chain cases when they 
are in the equilbrium domain structures. 
The last section consists of the conclusions and discussion. In the  
appendix, the details of the methods and mathematical tricks in the series 
calculations are given.

\section{Method}

In the continuum approximation, we found that
the vortex density increases at closer distances to the magnetic
domain walls. Based on this fact and the symmetry of the stripe domain
structure, it is reasonable to consider that the vortices and antivortices
form periodic structures on straight chains along the $y$ direction. Even
though it is not clear how many chains are associated with each domain, we
can still make progress
toward understanding discrete vortex lattices.
To this end,  
five stripe domain configurations
in which vortices are situated periodically on chains
 are proposed. From this point on, the configurations with $N$ vortex chains per 
stripe domain are labeled as $N$ state.

\begin{figure}
\centering
\subfigure[1st Case] 
{
    \label{fig:1stcase}
    \includegraphics[angle=0,width=3cm]{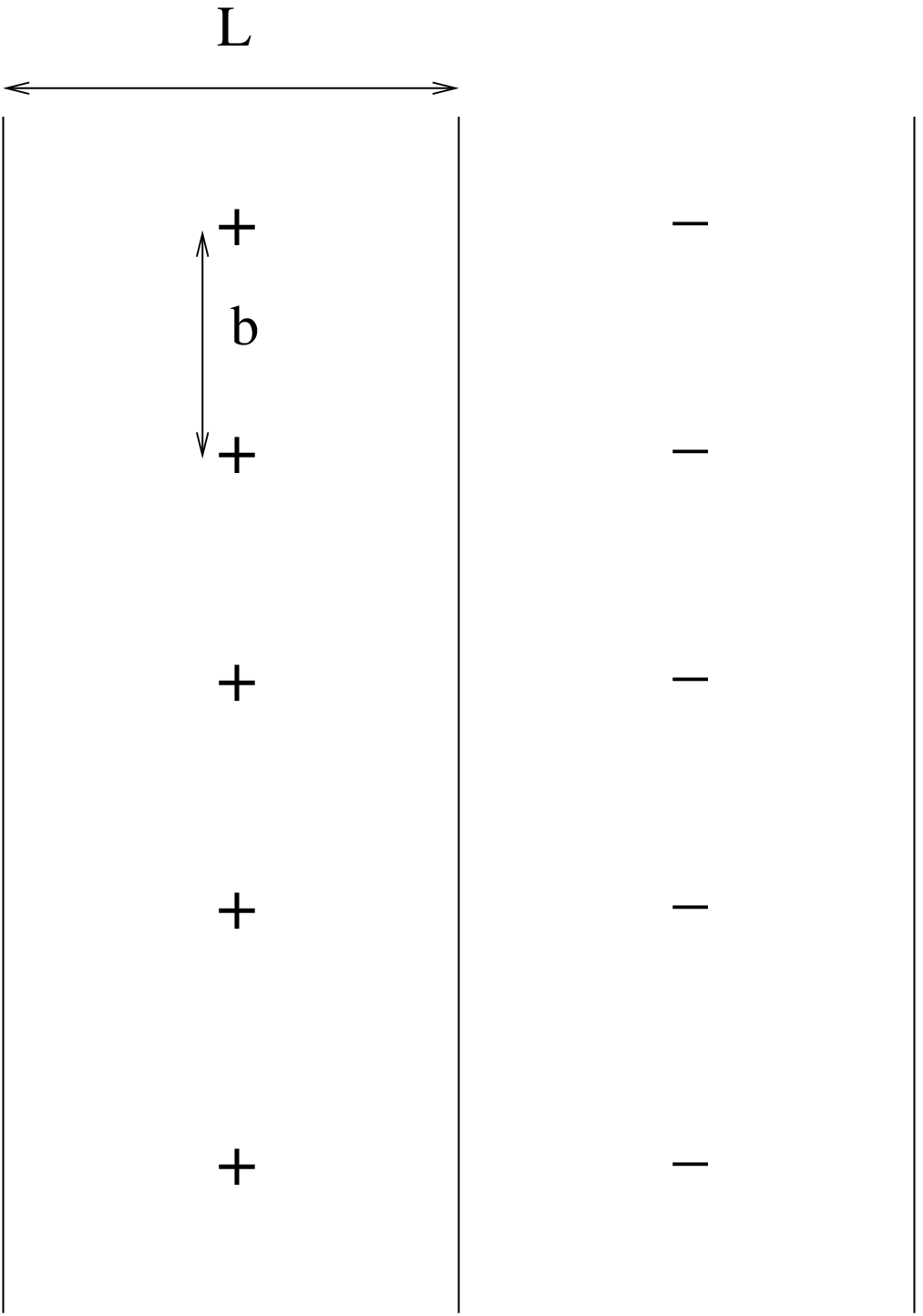}
}
\hspace{1cm}
\subfigure[2nd Case] 
{
    \label{fig:2ndcase}
    \includegraphics[angle=0,width=3cm]{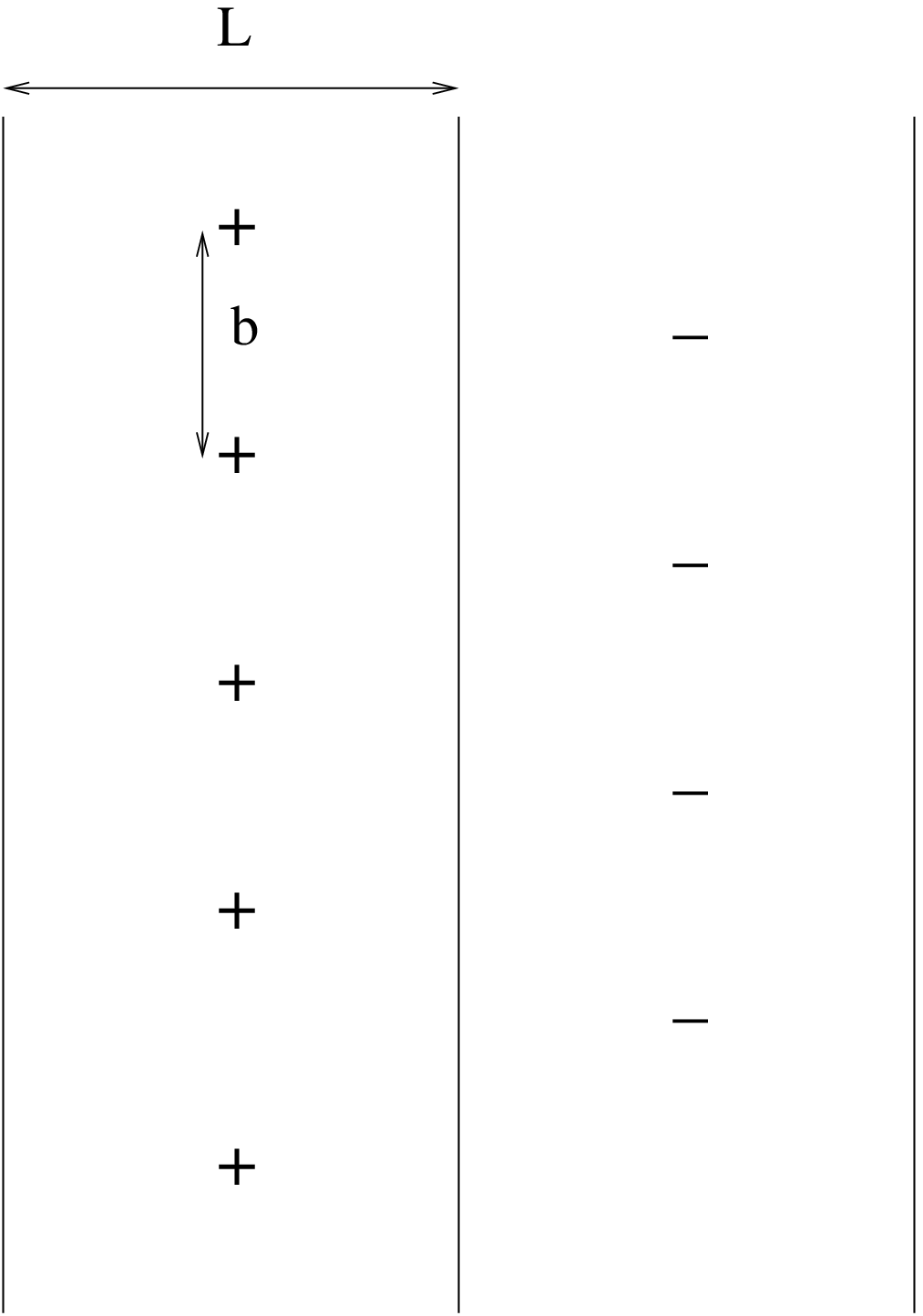}
}
\vspace{1cm}
\subfigure[3rd Case:1] 
{
    \label{fig:3rdcasea}
    \includegraphics[angle=0,width=3cm]{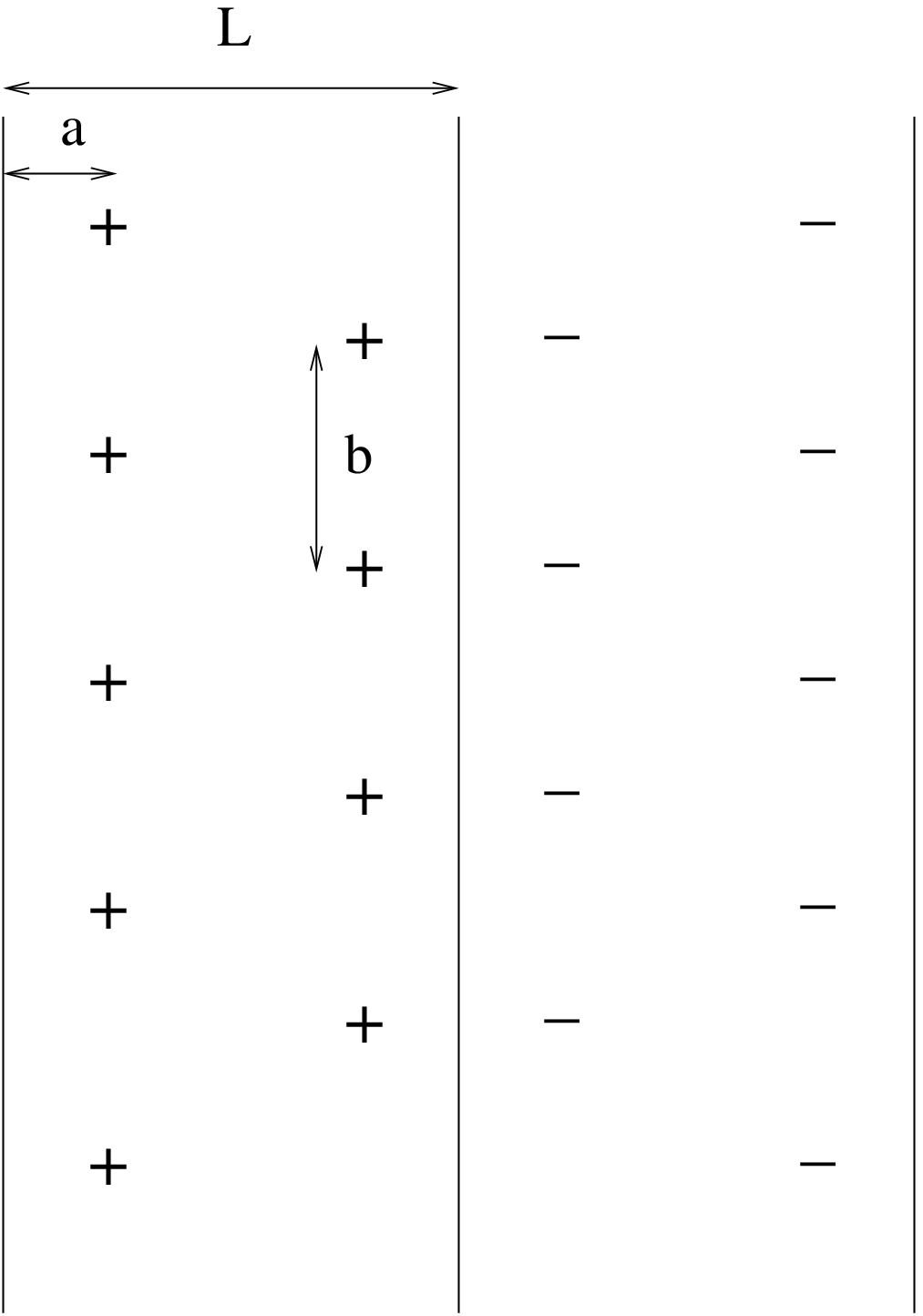}
}
\hspace{1cm}
\subfigure[3rd Case:2] 
{
    \label{fig:3rdcaseb}
    \includegraphics[angle=0,width=3cm]{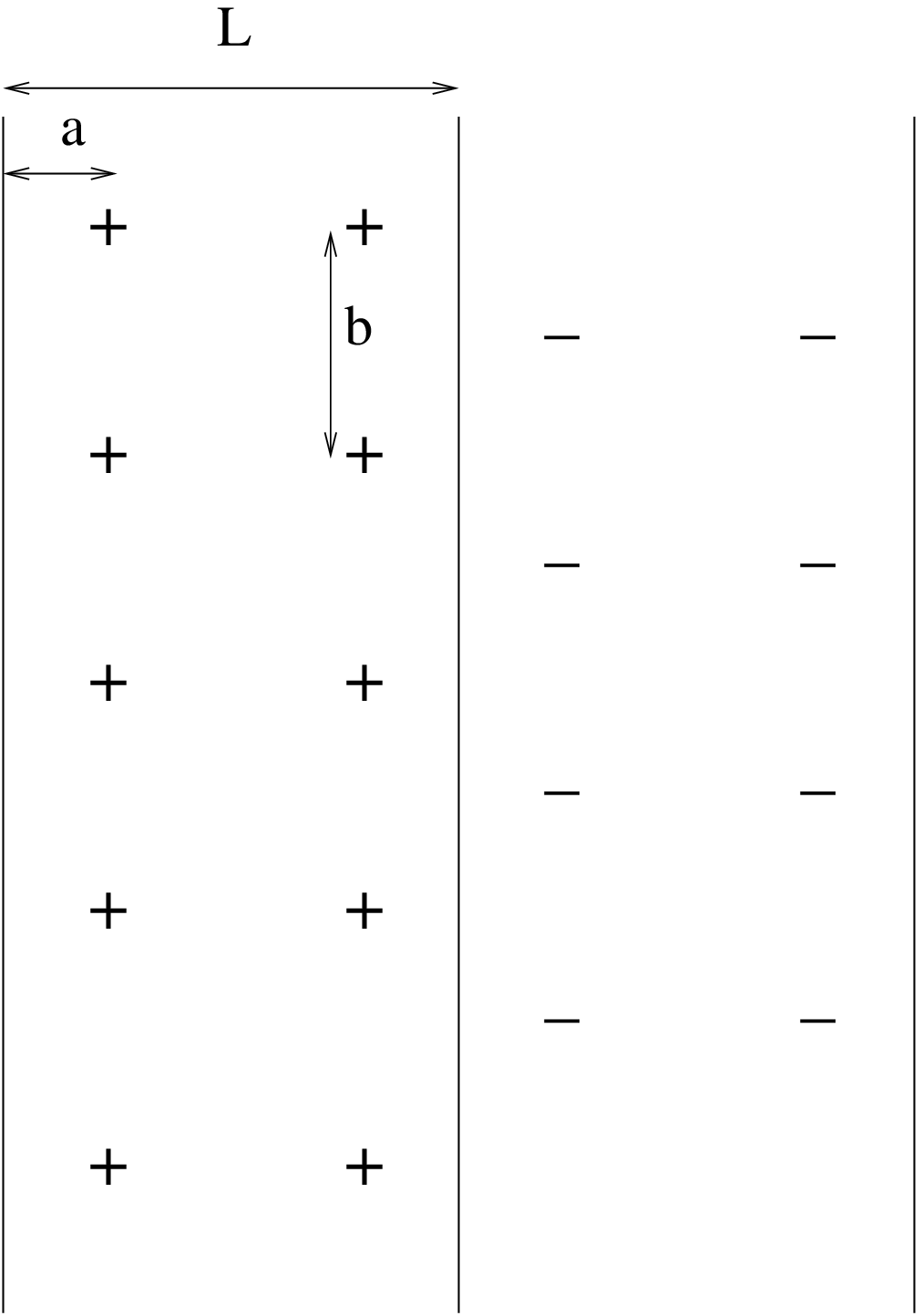}
}
\vspace{1cm}
\subfigure[4th Case] 
{
    \label{fig:4thcase}
    \includegraphics[angle=0,width=3cm]{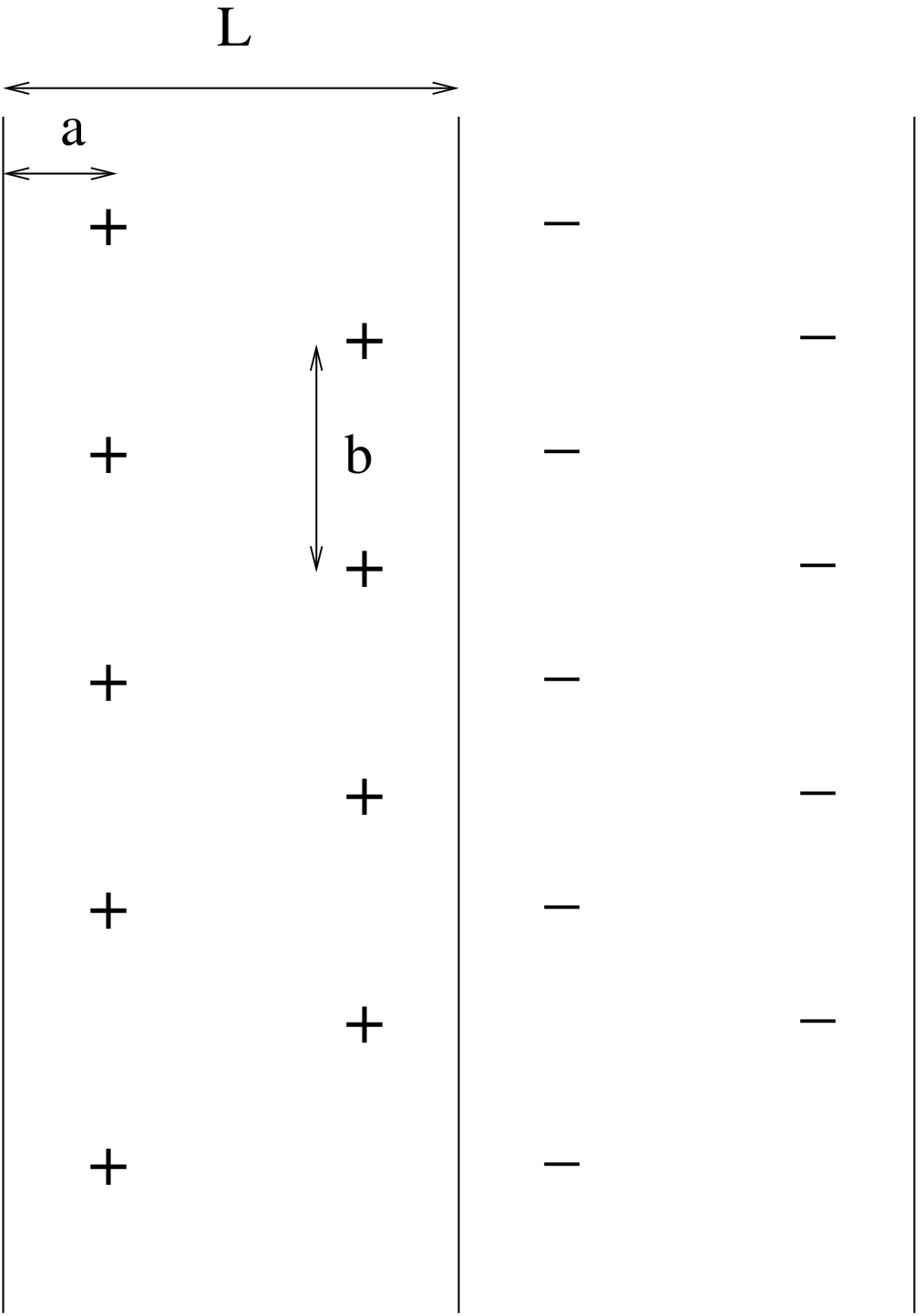}
}
\hspace{1cm}
\subfigure[5th Case] 
{
    \label{fig:5thcase}
    \includegraphics[angle=0,width=3cm]{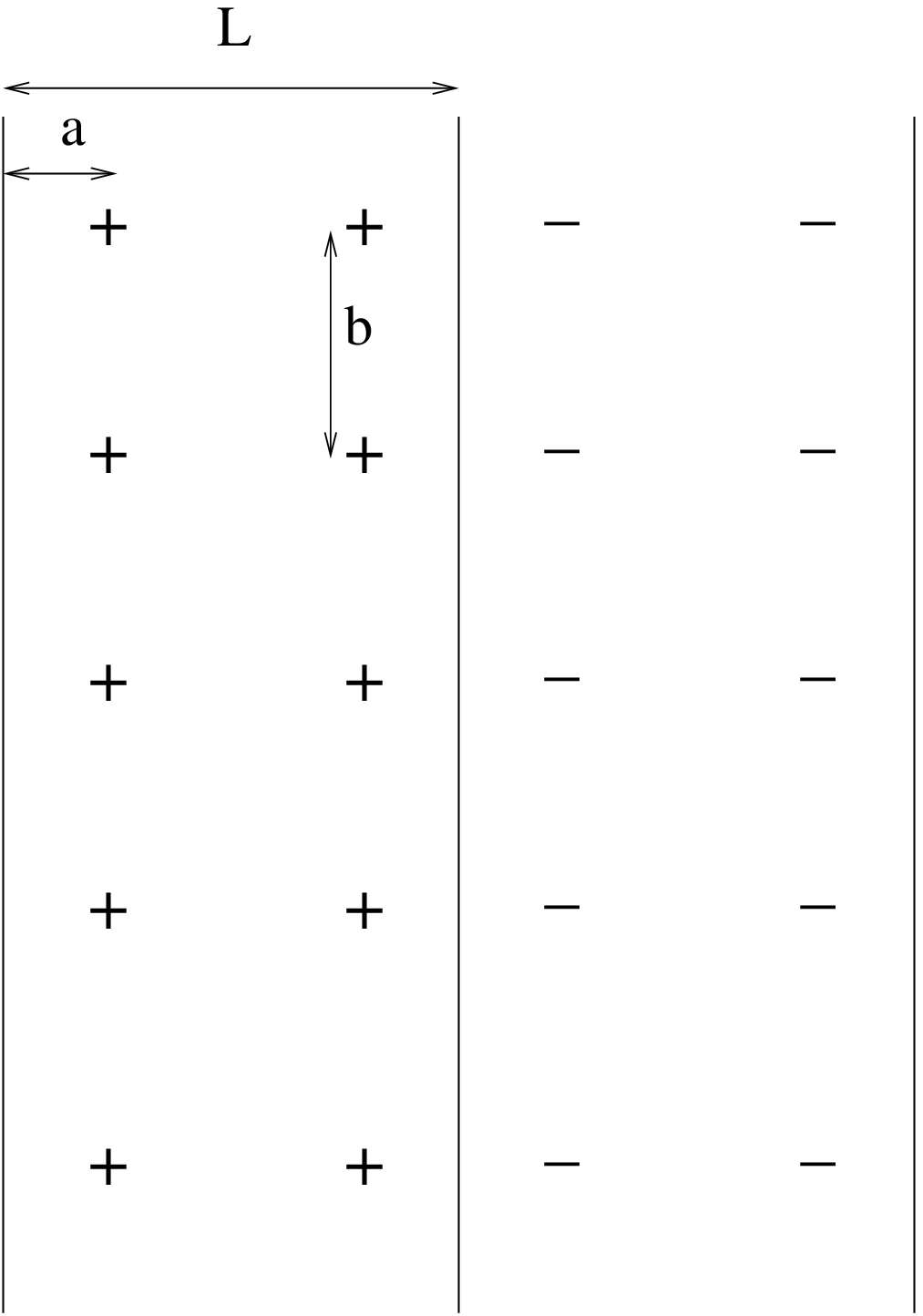}
}
\caption{Proposed configurations for $N=1$ and $N=2$ states.}
\label{fig:anal} 
\end{figure}

In two of the proposed cases, there is one chain per stripe ($N=1$ states), located in the
middle of the domain. In this case, two configurations of vortex lattice
are possible. First, the vortices and the
antivortices in a neighboring domains are alongside one another (see
Fig. \ref{fig:1stcase}); second, they are shifted by half period $b/2$ 
along the
$y$ direction, where $b$ is the distance between two nearest vortices on
the chain (see
Fig. \ref{fig:2ndcase}).


In the other three candidate lattice structures, there are two chains per
stripe domain $N=2$ states, at a distance $a$ from the magnetic domain walls. The
possible situations in the two-chain cases are as follows. First, chains
in the same domains are shifted by a half period $b/2$ along the $y$
direction (see
Fig. \ref{fig:3rdcasea}). However, the neighbor vortices and antivortices are next to
each other. Second, they are just shifted by a half period $b/2$ 
(see
Fig. \ref{fig:4thcase}). Third, the vortices and the antivortices are 
simply side by
side on the chains (see
Fig. \ref{fig:5thcase}). 


Our next step was to write the energies of these five candidates. To this
end, we used the energy  equations for periodic systems developed elsewhere 
\cite{ser-physica}.   


\begin{eqnarray}
u_{vv} &=& \frac{\phi_0^2}{4 \pi{\cal A}^2} \sum_{\bf G}
\frac{ |F_{\bf G}|^2}{G ( 1 + 2 \Lambda G)}, \label{puv} \\
u_{mv} &=& -
\frac{\phi_0}{\cal A} \sum_{\bf G} \frac{m_{z {\bf G}}F_{-\bf G}}{1 + 2
\Lambda G}, \label{pumv}\\
u_{mm} &=& -2 \pi \Lambda \sum_{\bf G} \frac{G^2
|{\bf m}_{z\bf G}|^2}{1 + 2 \Lambda G}. \label{pumm}
\end{eqnarray}

In these equations, the vortex
configurations differ by their form factors. We can obtain them from
$F_{\bf G} = \sum_{{\bf r}_i} n_i e^{i {\bf G} \cdot {\bf r}_i}$, where
the ${\bf G}$'s are the reciprocal vectors of the periodic structures, the
${\bf r}_i$ are the positions of the vortex centers, and $n_i$ are the
charge of the vortex. In our proposed models, ${\bf G} = ( ( 2 r + 1 )
\frac{\pi}{L}, 2 s \frac{\pi}{b})$ and $n_i = \pm 1$. Table \ref{tabF}
gives the form factors of each configuration in the order they are
described above.

\begin{table}[h] 
\caption{The form factors of vortices in the proposed
configurations. \label{tabF}} 
\begin{center} 
\begin{tabular}{|c|c|}\hline
{ configuration}&{$F_{\bf G}$} \\ \hline 
1 & $i (-1)^r$ \\ \hline 
2 & $i (-1)^r ( 1 + (-1)^s)$ \\ \hline 
3 & $2 i \sin G_x a ( 1 + (-1)^s)$ \\ \hline 
4 & $e^{i G_x a} - (-1)^ s e^{- i G_x a}$ \\ \hline 
5 & $ 2 i \sin G_x a$ \\ \hline 
\end{tabular} 
\end{center} 
\end{table}

\noindent Note that, in Table \ref{tabF}, the form factor for the third
configuration also belongs to the case in which the vortex and the
antivortex chains are shifted by half period only in the neighbour
domains, not in the same domain (see
Fig. \ref{fig:3rdcaseb}). Since information about the vortex
lattice is carried only by the form factors, there is no need to consider
the above-mentioned case separately.

In our calculations, the divergent part of the series must be extracted
carefully. We show below a detailed analysis of the series equations for
each candidate. We start with the self-interaction energy of the magnetic
layer $U_{mm}$, since it is the same for each configuration. For the
periodic structures, it is given by Eq. (\ref{pumm}). Direct substitution of
the Fourier coefficient of the stripe phase $m_{z{\bf G}}=\frac{2 i m
}{\pi( 2 r + 1)}$ into Eq. (\ref{pumm}) gives the self-interaction of the
magnetic layer per unit cell as

\begin{equation} 
u_{mm} = - \frac{8 m^2}{L} \sum_{r=0}^{\infty}
\frac{1}{\frac{L}{2 \pi \Lambda} + 2 r + 1}, \label{Hmmser} 
\end{equation}

\noindent where $\psi^{(0)} (x)$ is the polygamma function of zeroth order
\cite{abrom}.  In our numerical calculations, we write the logarithmic
term in Eq. (\ref{Hmmser}) as $\ln (\Lambda/l) + \ln (L/\Lambda)$ and then
incorporate the $-4m^2 \ln (\Lambda/l)$ term in the renormalized
$\varepsilon_{dw}^{ren}$. Another energy term with a divergent series is
the vortex energy, in general given by Eq. (\ref{puv}). The logarithmic
divergence in this term stems from the vortex self-energies. We first
split Eq. (\ref{puv})  into two parts as follows:

\begin{equation} 
u_{vv} = \frac{\pi \varepsilon_0}{2 L^2 b^2} \sum_{\bf
G}\left[ \frac{ |F_{\bf G}|^2}{G^2} - \frac{ |F_{\bf G}|^2}{G^2 (1 + 2
\Lambda G)}\right]. \label{vorener} 
\end{equation}

\noindent Note that the area of the unit cell is $2 L b$. The first
term of the series above contributes to the self-energies of the vortices;
whereas, the second term is the vortex-vortex energy and will be left in
the series form. For each form factor in Table \ref{tabF}, the series in
the first term can be transformed to the form of \\
$\sum_{r=-\infty}^{\infty}\sum_{s=-\infty}^{\infty} 1/((2 r + 1)^2 x^2 +
s^2)$, where $x$ is constant, and depends on the form factor. A detailed
analysis of such a series is given in the appendix.

The next step is to find the vortex energy and the interaction energy of
the magnetization and vortices
 for each configuration. In the calculation of $u_{mv}$, we take the
Fourier coefficient of the magnetization to be $\frac{4 i m }{( 2 r +
1)}\delta (G_y)$. The fact that the stripe is infinite along the $y$
direction results in the additional term $2 \pi \delta (G_y)$. However, it
does not play any role in the calculation of $u_{mm}$. For numerical
analysis, these energies must be expressed in terms of dimensionless
parameters. To this end, we define dimensionless variables $\tilde \Lambda
= \Lambda/L$ , $\tilde b = b/L$ and $\tilde \varepsilon_{dw}
 = \varepsilon_{dw}^{ren} \Lambda/\varepsilon_0$.  The total energy
$\tilde U$ is measured in units of $\varepsilon_0/\Lambda^2$. In addition,
we introduce the dimensionless magnetic energy as $\tilde U_{mm} =
u_{mm}/(\varepsilon_0/\Lambda^2)$. In terms of these parameters, the
energy of the first configuration reads


\begin{widetext}

\begin{equation} 
\tilde U^{(1)} = \frac{\tilde \Lambda^2}{4 \tilde b}
\left(\ln \left( \frac{4 \Lambda}{ e^C \tilde \Lambda \xi}\right) + 2 
f_v^{(1)} 
( \tilde
\Lambda) - \frac{2 f_{vv}^{(1)} ( \tilde \Lambda, \tilde b )}{ \tilde b
\pi} - \frac{16 m \phi_0}{\varepsilon_0} f_{mv}^{(1)} (\tilde \Lambda )
\right) + \tilde U_{mm} + \tilde \varepsilon_{dw} \tilde \Lambda,
\label{conf1} 
\end{equation}


\noindent where,

\begin{eqnarray} 
f_v^{(1)} &=& \sum_{r=0}^{\infty} \frac{\coth( ( 2 r + 1)
\frac{\tilde b \pi}{2}) -1}{2 r + 1}, \nonumber \\ 
f_{vv}^{(1)} &=&
\sum_{r,s = - \infty}^{\infty} \frac{1}{\left(( 2 r +1 )^2 + \frac{4
s^2}{\tilde b^2})(1+2 \pi \tilde \Lambda \sqrt{( 2 r +1 )^2 + \frac{4
s^2}{\tilde b^2}})\right)}, \nonumber \\
f_{mv}^{(1)} &=& \sum_{r=0}^{\infty} \frac{(-1)^r}{(2 r + 1 ) ( 1 + 2 \pi
\tilde \Lambda( 2 r + 1 ))}. \label{fmmfmv} 
\end{eqnarray}

\end{widetext}


The form factor for the second configuration survives only for even values
of $s$. Then, the dimensionless energy of the second configuration is
found to be


\begin{widetext}

\begin{equation} 
\tilde U^{(2)} = \frac{\tilde \Lambda^2}{2\tilde b}
\left(\ln \left( \frac{4 \Lambda}{e^C \tilde \Lambda \xi}\right) + 
2f_v^{(2)} ( \tilde
\Lambda) - \frac{4 f_{vv}^{(2)} ( \tilde \Lambda, \tilde b ) }{\tilde b
\pi} - \frac{16 m \phi_0}{\varepsilon_0} f_{mv}^{(2)} (\tilde \Lambda
)\right) + \tilde U_{mm} + \tilde \varepsilon_{dw} \tilde \Lambda,
\label{conf2} 
\end{equation} 
\noindent where $f_{mv}^{(2)} = f_{mv}^{(1)}$ and,


\begin{eqnarray} 
f_v^{(2)} &=& \sum_{r=0}^{\infty} \frac{\coth (( 2 r + 1
)\frac{\tilde b\pi }{4}) -1}{2 r + 1}, \nonumber \\
f_{vv}^{(2)} &=& \sum_{r,s = - \infty}^{\infty} \frac{1}{\left(( 2 r +1 )^2 +
\frac{16 s^2}{\tilde b^2})(1+2 \pi \tilde \Lambda \sqrt{( 2 r +1 )^2 +
\frac{16 s^2}{\tilde b^2}})\right)}. \label{fvfmv2} 
\end{eqnarray}

\end{widetext}


In the third configuration, as in the second configuration, only even
values of $s$ contribute to the energy. In the first two configurations,
the square of their form factors enter the vortex energy as a constant.
However, in this case, the square of the sine function appears. In the 
appendix, the calculation of  the series in the presence of
such functions is shown. Introducing the dimensionless parameter $\tilde a = a/L$,
the energy functional of the third configuration becomes


\begin{widetext}

\begin{eqnarray} 
\tilde U^{(3)} &=& \frac{\tilde \Lambda^2}{\tilde b}
\left( \ln \left( \frac{4 \Lambda}{e^C \tilde \Lambda \xi}\right)-\ln 
(\cot(\pi \tilde a))
+ 4 f_v^{(3)} ( \tilde \Lambda, \tilde a) - \frac{8 }{\tilde b \pi}
f_{vv}^{(3)} ( \tilde \Lambda, \tilde a, \tilde b ) - \frac{16m
\phi_0}{\varepsilon_0} f_{mv}^{(3)} (\tilde \Lambda,\tilde a )\right) 
\nonumber \\
&+& \tilde U_{mm} + \tilde \varepsilon_{dw} \tilde \Lambda, \label{conf3}
\end{eqnarray}


\noindent where,

\begin{eqnarray} 
f_v^{(3)} &=& \sum_{r=0}^{\infty} \frac{\coth (( 2 r + 1
)\frac{\tilde b \pi}{4} ) -1}{2 r + 1} \sin^2 (( 2 r + 1 ) \pi \tilde a),
\nonumber \\
f_{vv}^{(3)} &=& \sum_{r,s = - \infty}^{\infty} \frac{\sin^2 (( 2 r + 1 )
\pi \tilde a) }{\left(( 2 r +1 )^2 + \frac{ 16 s^2}{\tilde b^2})(1+ 2 \pi
\tilde \Lambda \sqrt{( 2 r +1 )^2 + \frac{16 s^2}{\tilde b^2}})\right)},
\nonumber \\
f_{mv}^{(3)} &=& \sum_{r=0}^{\infty} \frac{\sin (( 2 r + 1 ) \pi \tilde
a)}{(2 r + 1 ) (1+2 \pi \tilde \Lambda ( 2 r + 1 ))}. \label{fvfmv3}
\end{eqnarray}


\end{widetext}

 In the fourth configuration, the square of the form factor is: \\
$|F_{\bf G}|^2 = 2 - 2 (-1)^s \cos ( ( 2 r + 1 ) \pi \tilde a )$. Even and
odd values of $s$ give different contributions. Then, we can calculate the
vortex energy for even $s$ and odd $s$ separately. Employing similar
techniques, we find
 

\begin{widetext}

\begin{equation} 
\tilde U^{(4)} = \frac{\tilde \Lambda^2}{2 \tilde b}
\left( \ln \left( \frac{4 \Lambda}{e^C \tilde \Lambda \xi}\right) + 2 
f_v^{(4)} ( \tilde
\Lambda, \tilde a) - \frac{4}{\tilde b \pi} f_{vv}^{(4)} ( \tilde \Lambda,
\tilde a, \tilde b ) - \frac{16m \phi_0}{\varepsilon_0} f_{mv}^{(4)}
(\tilde \Lambda,\tilde a )\right) + \tilde U_{mm} + \tilde 
\varepsilon_{dw} \tilde
\Lambda, \label{conf4} 
\end{equation}


\noindent where $f_{mv}^{(4)} = f_{mv}^{(3)}$ and,

\begin{eqnarray} 
f_v^{(4)} &=& \sum_{r=0}^{\infty} \frac{\coth (( 2 r + 1
)\frac{\pi \tilde b}{4} ) -1}{2 r + 1} \sin^2 (( 2 r + 1 ) \pi \tilde a)
+ \sum_{r=0}^{\infty} \frac{\tanh (( 2 r + 1 )\frac{\pi \tilde b}{4} )
-1}{2 r + 1} \cos^2 (( 2 r + 1 ) \pi \tilde a), \nonumber \\ 
f_{vv}^{(4)}
&=& \sum_{r,s = - \infty}^{\infty} \frac{\sin^2 (( 2 r + 1 ) \pi \tilde
a)}{\left(( 2 r +1 )^2 + \frac{16 s^2}{\tilde b^2})(1+ 2 \pi \tilde 
\Lambda
\sqrt{( 2 r +1 )^2 + \frac{16 s^2 }{\tilde b^2}})\right)} \nonumber \\
 &+& \sum_{r,s = - \infty}^{\infty}\frac{\cos^2 (( 2 r + 1 ) \pi \tilde
a)}{ \left(( 2 r +1 )^2 + \frac{4 ( 2 s+1 )^2}{\tilde b^2}) (1+ 2 \pi 
\tilde \Lambda
\sqrt{( 2 r +1 )^2 + \frac{4 (2 s +1)^2 }{\tilde b^2}})\right)}. 
\end{eqnarray}

\end{widetext}


The form factor for the fifth case resembles that of the third case with
an exception. That is, in the third case, only even values of $s$ are
taken into account, while all integers contribute to the sum over $s$ in
the fifth case. Keeping this in mind, obtaining the 
dimensionless energy for the last case is straightforward:


\begin{widetext}

\begin{eqnarray} 
\tilde U^{(5)} &=& \frac{\tilde \Lambda^2}{2 \tilde b}
\left( \ln \left( \frac{4 \Lambda}{e^C \tilde \Lambda \xi} \right) -\ln ( 
\cot(\pi \tilde a)) + 4 f_v^{(5)} ( \tilde \Lambda, \tilde a) - \frac{4}{\tilde b \pi}
f_{vv}^{(5)} ( \tilde \Lambda, \tilde a, \tilde b ) - \frac{16 m
\phi_0}{\varepsilon_0} f_{mv}^{(5)} (\tilde \Lambda ,\tilde a)\right) 
\nonumber \\
&+& \tilde U_{mm} + \tilde \varepsilon_{dw} \tilde \Lambda, \label{conf5}
\end{eqnarray}


\noindent where $f_{mv}^{(5)} = f_{mv}^{(3)}$ and, 
\begin{eqnarray}
f_v^{(5)} &=& \sum_{r=0}^{\infty} \frac{\coth (( 2 r + 1 )\frac{\pi \tilde
b}{2} ) -1}{2 r + 1} \sin^2 (( 2 r + 1 ) \pi \tilde a), \nonumber \\
 f_{vv}^{(5)} &=& \sum_{r,s = - \infty}^{\infty} \frac{ \sin^2 (( 2 r + 1
) \pi \tilde a)}{\left(( 2 r +1 )^2 + \frac{4 s^2}{\tilde b^2})(1+ 2 \pi \tilde
\Lambda \sqrt{( 2 r +1 )^2 + \frac{4 s^2}{\tilde b^2}})\right)}. 
\label{fvfmv5}
\end{eqnarray}

\end{widetext}


\section{Computational Methodology}

The series in
$f_v^{(i)}$ converges very fast when $r_{max} > 200$, while the series in
$f_{vv}^{(i)}$ converges rather slowly. When
$r_{max}>4000$ and
$s_{max} >4000$, the results do not change up to the 6th decimal point in the
energy,  where $i$ labels the particular domain configuration.
To make sure of this accuracy in the calculations,
we take $r_{max}= 600$ for $f_v^{(i)}$,  $r_{max}= 5000$
for $f_{mv}^{(i)}$,  and $r_{max}= 5000$ and
$s_{max} = 5000$ for  $f_{vv}^{(i)}$ in Eqs.(\ref{conf1}, \ref{conf2},
\ref{conf3}, \ref{conf4}, \ref{conf5}). In addition, the respective error
deviations for
$\Lambda/L$,$b/L$ and $a/L$ in numerical calculations are $\pm 0.005$,$\pm
0.0005$ and $\pm 0.0125$. In the numerical computations   of Eqs.(\ref{conf1}, \ref{conf2},
\ref{conf3}, \ref{conf4}, \ref{conf5}), we take $\ln (4 \Lambda/(e^C \xi )) = 5.57$.

\section{The Vortex Chain States in Domains with Fixed Period}

\begin{figure}[t]
\begin{center}
\includegraphics[angle=270,width=3.5in,totalheight=3.5in]{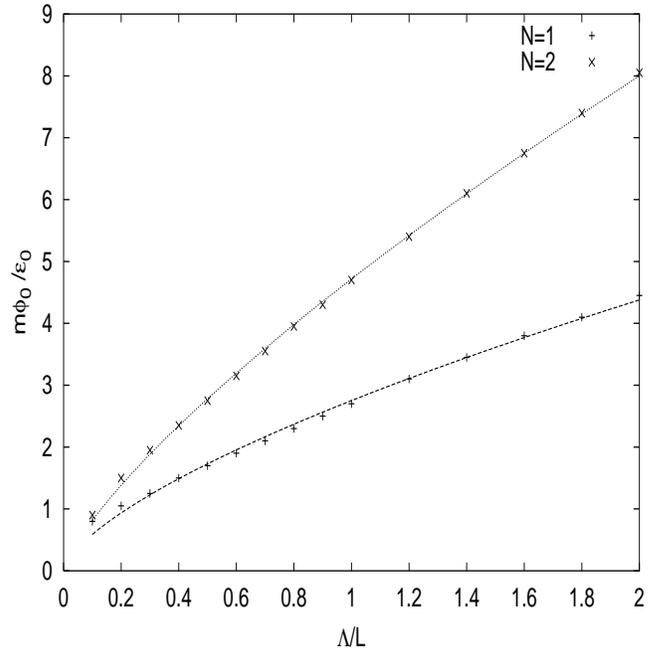}
\caption{Phase diagram showing transition from the Meissner state to the 
mixed state, depending on $m \phi_0/\varepsilon_0$ and $\Lambda/L$. The 
region below the $N=1$ curve corresponds to  the Meissner state, while 
$N=1$ state exists in the 
area 
between the $N=1$ and $N=2$ curves. The $N=2$ state becomes energetically 
favorable above the $N=2$ curve.   
\label{meissner}}
\end{center}
\end{figure}

Here, the conditions for the transition from the 
Meissner state to the mixed state when the vortices first spontaneously 
appear, are determined. As described in the second section, the total energy of the bilayer consists of 
the self-vortex energies $u_v$, vortex-vortex interaction $u_{vv}$, 
vortex-magnetization 
interaction $u_{mv}$, the self-interaction of magnetic domains $u_{mm}$ 
and 
the domain wall energy $u_{dw}$. The last two terms are necessary only to 
determine the equilibrium domain size. Therefore, they are irrelevant when 
the domain period is fixed, and they will be ignored in the further  
calculations to determine when the Meissner state-mixed state transition 
occurs. Now, we call the total energy  of interest, the effective 
energy to avoid any confusion in the rest of the paper. The effective energy 
 has the form $U= u_{v} + u_{vv} + u_{mv}$. The detailed versions of these terms 
for each proposed configuration were described in the second section. Next, we calculate 
the equilibrium effective energies of  each configuration for fixed values 
of $m \phi_0 / \varepsilon_0$ and $\Lambda/L$. In doing so, the effective 
energies are  minimized with respect to the vortex positions. 
The necessary condition for the vortices to appear spontaneously is 
$U_{eff} < 0$. Our numerical calculations show that the configurations 
with a single vortex chain per domain ($N=1$ state ) appear first in the 
mixed state.  Note that there are  two configurations for the $N=1$ state. In our 
calculations, it turns out that they are indistinguishable when the 
transition occurs. Next, the values of $m \phi_0/\varepsilon_0$ that 
make $U_{eff}^{N=1}$  zero  for various  values of the domain period, 
are calculated. Thus, the curve that separates the Meissner state and 
$N=1$ state is 
obtained (see the bottom curve in Fig. \ref{meissner}). The fit of  our 
numerical data yields an equation for this curve as: 
$m \phi_0/\varepsilon_0 = 2.75 (\Lambda/L)^{0.67}$. Earlier,
Laiho {\it 
et al.} \cite{laiho}
calculated  critical  magnetizations for two cases: penetration of a 
vortex
from a domain center and penetration of a vortex semiloop from a domain wall in a FSB, in which  both  FM and SC
layers are thick.  For the former case, they found that the critical 
magnetization decreases with the increase of
domain period. This result is in full agreement with ours. When vortices 
appear first in FSB, the inter-vortex distance $b$  grows. With 
further increase of the 
magnetization, they get close. Note that our results  here
do not give any information regarding
which configuration of $N=1$  becomes favored in the mixed
state.

  We also studied the transition from the $N=1$ state to the $N=2$ state. The 
necessary conditions for this transition are $U_{eff}^{N=2} < U_{eff}^{N=1}$ 
and $U_{eff}^{N=2} < 0$. Our calculations show that the third 
configuration wins over the fourth and fifth configurations for the $N=2$ 
state. Note that the effective energy of the third configuration corresponds to 
two cases that have the same structure factor. Which one is likely to win over 
is discussed in the next section.  
Following steps similar to those described above, we 
obtained the curve   $m \phi_0/\varepsilon_0  = 4.72 (\Lambda/L)^{0.76}$ 
(see the top curve in Fig. \ref{meissner}). Furthermore, when the $N=2$ state 
wins over the $N=1$ state, $b$ and $a$ are on the order of  a few $\Lambda$. 
The 
increasing magnetization leads to a significant decrease of these 
parameters.

\section{The Vortex Chain States in  the Equilibrium Domain Structures}
\begin{figure}[t]
\begin{center}
\includegraphics[angle=270,width=3.5in,totalheight=3.5in]{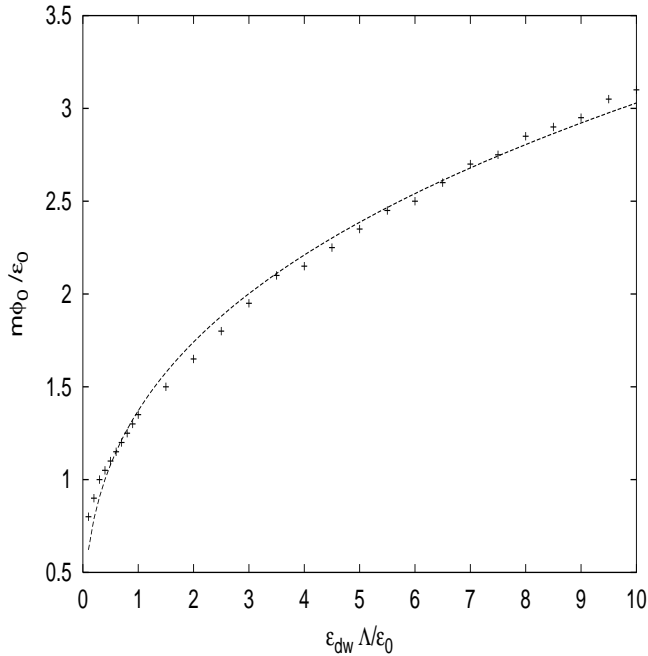}
\caption{The phase diagram for the third case  $N=2$ state. The $N=2$ state becomes stable in the region above the curve. 
\label{phasem1}} \end{center}
\end{figure}

Here, we first investigated when the proposed cases become energetically favorable. To this end, 
we checked when the equilibrium energies of the cases first become negative.  
To do so, 
the energies of five proposed cases 
Eqs.(\ref{conf1},\ref{conf2},\ref{conf3},\ref{conf4},\ref{conf5}) for 
different values of $m \phi_0/\varepsilon_0$ and $\tilde 
\varepsilon_{dw}$ were calculated by minimizing their energies with 
respect to 
the domain size and the vortex positions. From this procedure, one can determine which case's equilibrium energy becomes negative first and 
then calculate the corresponding values of $m \phi_0/\varepsilon_0$ and $\tilde \varepsilon_{dw}$. 
Our calculations show that the equilibrium energy  of the third case 
associated with the $N=2$ state turns to negative first. When 
the $N=2$ state
first appears, the domain size and inter vortex distance on the same chain 
are on the order of a few tens of $\Lambda$.
This suggests that, 
in equilibrium, vortex chain states appear near the domain walls. It is expected that vortex chain states proliferate
with further increase of the magnetization. This result is consistent with our 
previous work in which the vortex density in 
the 
continuum aproximation 
increases substantially near the magnetic domain walls. In Fig. \ref{phasem1}, the phase diagram for the $N=2$ state is depicted. The dots in the 
figure are 
obtained from our simualtions. The curve is the fit to the simulation data. The fit gives
the curve $m \phi_0 / \varepsilon_0 = 1.37 {\tilde \varepsilon_{dw}}^{0.34}$. Below the curve, there is not any stable configuration, 
while, above the curve, 
the third configuration corresponding to $N=2$ state exists. 
   
Nonetheless, this information is not  enough for us to
understand the equilibrium structure, since
 $\tilde U^{(3)}$  corresponds to two different
configurations with the same structure factor. At this point, further
analysis is neededto determine
which configuration  is more likely.  To this end,
we calculated the minimal energy of each
case in both continuum and discrete regimes, depending on  
 $m \phi_0/\varepsilon_0$ at fixed $\tilde \varepsilon_{dw}$.
 The equilibrium energies for these cases 
in discrete and continuum regimes are given in Table \ref{energies}.
Note that small and large values of $m \phi_0/\varepsilon_0$ for each $\tilde \varepsilon_{dw}$ correspond
to continuum and discrete regimes, respectively.  

\begin{widetext}

\begin{table}[h]
\caption{Equilibrum energies for proposed 
configurations. Two columns on the left are input. \label{energies}}
\begin{center}
\begin{tabular}{|c|c|c|c|c|c|c|}\hline
{$\tilde \varepsilon_{dw}$}&{$m\phi_0/\varepsilon_v$}&{$\tilde 
U_{1}$}&
{$\tilde U_{2}$}&{$\tilde U_{3}$}&{$\tilde U_{4}$}&{$\tilde U_{5}$}\\ 
\hline
 0.01 & 5 & -2.58454176 & -2.58455668 & -3.36407195 & -3.36404625 & -3.36404615  \\ \hline
 0.01 & 20 & -65.98296440 & -65.98296500 & -89.20105311 & -89.20030961 &-89.20030943  \\ \hline
 0.1 & 5 & -2.54211623 & -2.54211637 & -3.33057991 & -3.33054828  & -3.32949063 \\ \hline
 0.1 & 20 & -65.87136440 & -65.87136500 & -89.10475311 &-89.10310961 &-89.10310943  \\ \hline
 1  & 5  & -2.16635495 & -2.16637343 & -3.01972350 &-3.01972355 &-3.01972347 \\ \hline
 1  & 20 & -64.78187766 & -64.781878054 & -88.15826391 & -88.15826394 &-88.15826367  \\ \hline
10 & 5 & -0.35644861 & -0.35667498 & -1.21258278 & -1.21232397& -1.21232396\\ \hline 
10 &  25 & -95.00241100 & -95.00247780 & -134.28090345 & -134.27589780& -134.27589769\\ \hline
\end{tabular}
\end{center}
\end{table}


\end{widetext}

In our numerical calculations,  we find that all proposed configurations are stable in both  
the discrete and continuum regimes, indicating that our method works well 
in both regimes. As in the previous section, the third case associated with the $N=2$ 
state wins over 
the other two chains-per-domain configurations. As said before, the energy of the third case corresponds to two configurations.
Which configuration is more likely to appear in equilibrium can be figured out 
from simple 
physical considerations. Namely, in FSB, the equilibrium structure is 
determined by the competition between vortex-vortex  and vortex-magnetization interactions.  
The former favors vortices and antivortices in the neighbor domains  
to line up in a transverse direction (perpendicular to the magnetic domain wall), whereas the latter prefers vortices and antivortices to be shifted so 
that gain in energy is maximized. When vortices are next
to each other on either side of the magnetic domain wall, the magnetic
fields they produce cancel out each other. From the numerical results, it 
is obvious that vortex-magnetization interaction wins the competition and 
results in half-way shifting of vortices, if one compares the energies of 
the 1st and 2nd cases. Then, vortex-magnetization interaction is the dominant 
factor.
By the same token, one can understand what is going on in double-vortex 
chain 
configurations. For instance, in the fifth configuration, energy gain due 
to 
vortex-magnetization interaction is diminished, since all the vortices are 
side by side. This explains why the equilibrium energy of the fifth 
configuration is higher than those of the third and fourth cases.
 In the fourth configuration, the vortices and 
antivortices in the neighbor domains are shifted half-way, so that this 
configuration must be preferred over the one in which they sit side by 
side in the neighbor domains according to the above arguments. However, 
an alternative configuration for the third case has two chains shifted half-way
in the neighbor domains instead of one chain as in the fourth case. 
Therefore, one 
might expect the gain to be even more than that in the fourth configuration.   
Another interesting result is that the system does not favor the $N=1$ state 
at all. 
Actually, this does not surprise us, since, in the
continuum approximation, we found that the vortex density increases near
the magnetic domain walls. This fact already suggests that the system
favors vortex chains being near the magnetic domain walls rather than a
single chain in the middle of the domain. 

\begin{figure}
\centering
\includegraphics[angle=270,width=7cm]{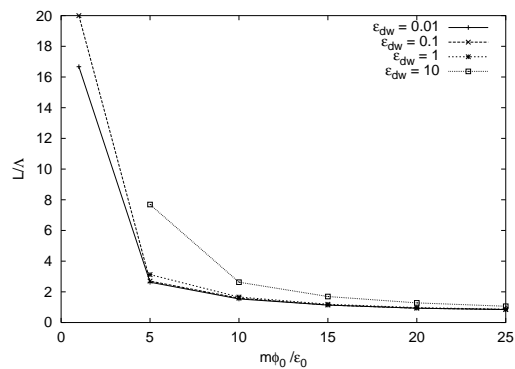}
\caption{$L/\Lambda$ versus $m \phi_0/\varepsilon_0$ for the third configuration.}
\label{fig:domainsize} 
\end{figure}
\begin{figure}
\centering
\includegraphics[angle=270,width=7cm]{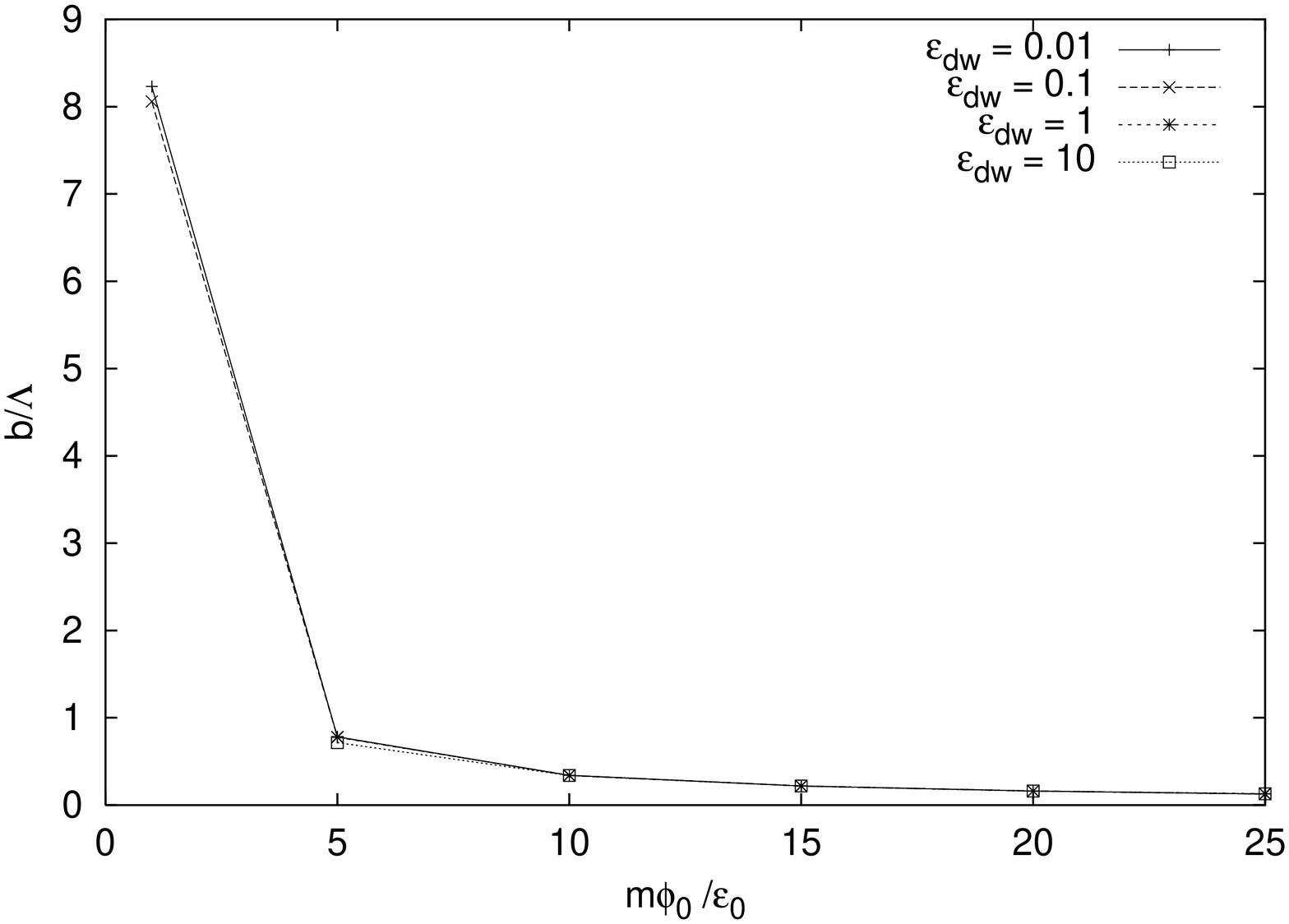}
\caption{$b/\Lambda$ versus $m \phi_0/\varepsilon_0$ for the third configuration.}
\label{fig:vvdist} 
\end{figure}
\begin{figure}
\centering
    \includegraphics[angle=270,width=7cm]{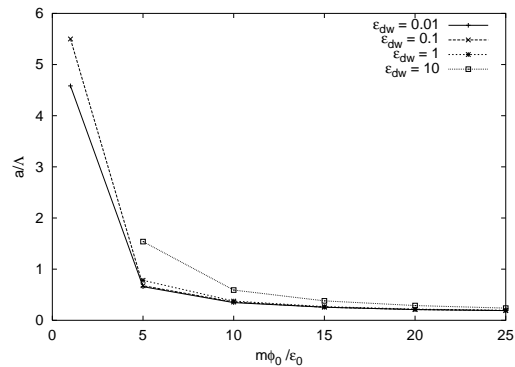}
\caption{$a/\Lambda$ versus $m \phi_0/\varepsilon_0$ for the third configuration.}
\label{fig:vmdist} 
\end{figure}

In numerical calculations, equilibrium domain size $L/\Lambda$,
vortex-vortex
distances  on the same chain $b/\Lambda$ and
vortex-magnetic domain wall distances $a/\Lambda$ are also calculated. 
Results for  the third configuration at 
various values of $m \phi_0/\varepsilon_0$  and $\tilde 
\varepsilon_{dw}$  are
depicted  in Figs. \ref{fig:domainsize}, \ref{fig:vvdist}, \ref{fig:vmdist}. Results for other configurations 
are not shown here, since they look quite similar.   
At fixed $\tilde \varepsilon_{dw}$, the further increase of $m \phi_0/\varepsilon_0$
shrinks the domain width, while a higher domain wall energy favors a larger domain width
at fixed $m \phi_0/\varepsilon_0$, as in usual ferromagnets. That is to say, ferromagnet
favors narrower domains to minimize the demagnetization energy, whereas domain wall energy
makes domains wider. The competition between these two energies determines the domain
size. Here, the parameter $m \phi_0/\varepsilon_0$ plays the role of
demagnetization energy.
Domain wall energy does not affect the distance between the
vortices located on the same chain. However, at larger values of $m \phi_0/\varepsilon_0$, 
the vortices on the same chain get closer. This implies that the
unit  cell area shrinks, and consequently vortex density per area increases.



\section{conclusions}

In this article, we reported our results on the lattices of discrete 
vortices 
in stripe domains in FSB using a method based on Maxwell-London 
equations. 
If $\varepsilon_{dw}\leq 4\tilde{m}^2$, the continuum approximation 
becomes invalid. Instead, we considered the discrete lattice of vortices
in which  the vortices were considered to be situated on
chains directed along the stripes. We analyzed
the vortex configurations up to two vortex chains in two cases: first, we 
took the domain period to be fixed, second, the domain period was not 
fixed. In the former case, we calculated the threshold magnetization at 
which the transition from the Meissner and mixed state occurred. We
showed that configurations corresponding to the $N=1$ state appeared first in 
the mixed state. With a further increase of magnetization, 
the $N=2$ state becomes more favorable. The critical magnetization for 
transition from the $N=1$ 
to the $N=2$ state was also calculated. In the second case, 
depending on the magnetization and the
magnetic domain wall energy, the equilibrium energy, the positions of the 
vortices and the
equilbrium
domain
size were calculated. 
According to our calculations, in equilibrium, vortices
on the either side of the magnetic domain walls are not side by side on
the
chains; instead,  they are shifted by a half period
along the stripe, while they are side by side in the same domain. The 
threshold magnetization as a function of domain wall energy for this case 
was calculated.

In numerical calculations, we also found that the vortex lattice is stable
for $\varepsilon_{dw} >4 \tilde m^2$. At this point, the domain size is
noticably larger than the effective penetration depth $\Lambda$, so the
continuum approximation is valid. Therefore, we
expect that the domain nucleation starts in the continuum
regime. This problem is left for the future research.  At constant $\tilde \varepsilon_{dw}$, with
increasing $m \phi_0 / \varepsilon_v$, the equilibrium size of the domain
decreases. In addition, the vortices on the chain get closer to each
other. These results agree with those obtained in the continuum
approximation.
As $\varepsilon_{dw}/4\tilde{m}^2$
increases, we expect that new vortex chains develop within the domains. We
leave the detailed analysis of this problem to another publication. In 
concluding, we believe that confined geometries, such as domains in 
FSB, lead to novel vortex structures, which might be 
experimentally 
investigated by means of scanning tunnelling spectroscopy.

\section{Acknowledgments}
The most of this work was done during my stay at the University of
Minnesota
and was
partially supported by the U.S. Department of Energy, Office of Science,
under Contract No. W-31-109-ENG-38.


\appendix*

\section{Calculations of  Series }




In this appendix, the detailed analysis of the series is given. First, the
series in the energy calculations of the periodic systems are analyzed;
second, the detailed calculation of the vortex density is shown. The
series we encounter in the energy calculations fall into two categories.
In the  first category, we sum over one variable. The series
in this category are in the form of $\sum_{r=1}^{r_{max}} 1/r$. Employing
the Euler-Maclaurin summation formula \cite{arfken}, the summation is
found with logarithmic accuracy as

\begin{equation} 
\sum_{r=1}^{r_{max}} \frac{1}{r} \approx \ln r_{max} + C,
\label{s1} 
\end{equation}

\noindent where $C \sim 0.577$ is the Euler-Mascheroni constant. If the
summation is performed over only odd integers, we can still transform our
series to Eq. (\ref{s1}). Namely,

\begin{eqnarray} 
\sum_{r=0}^{r_{max}} \frac{1}{2 r + 1} &\approx&
\sum_{r=1}^{2 r_{max} +1} \frac{1}{r} - \frac{1}{2} \sum_{r=1}^{r_{max}/2}
\frac{1}{r}, \nonumber \\ 
&\approx& \ln ( 2 r_{max} + 1 ) + C - \ln
(\frac{r_{max}}{2}) - \frac{C}{2}, \nonumber \\ 
&\approx& \frac{1}{2} ( \ln r_{max}
+ C + 2 \ln 2 ). \label{s2new} 
\end{eqnarray}

The other double series of interest here is in the form of

\begin{equation} 
I ( x ) = \sum_{r= -\infty}^{r=\infty} \sum_{s=
-\infty}^{s=\infty} \frac{1}{x^2 r^2 + s^2}, \label{s4new} 
\end{equation}

\noindent where $x$ is an arbitrary constant. Although Eq. (\ref{s4new}) is
logarithmically divergent, the sum over one of the variables can be done
easily. To this end, we perform the sum over $s$ first. In doing so, Eq.  
(\ref{s4new}) becomes $(2 \pi/x) \sum_{r=1}^{\infty} \coth (\pi x r)/r$
\cite{series}. This series is logarithmically divergent. In order to get
the logarithmic term , we add and subtract $1/r$. Using the result in
Eq. (\ref{s1}), finally we get

\begin{equation} 
I ( x ) \approx \frac{2 \pi}{x} \left[\sum_{r =1}^\infty
\frac{\coth ( \pi x r) - 1}{r} + \ln r_{max} + C\right]. \label{s5}
\end{equation}

\noindent Employing the same techniques, we give the results of the
different versions of Eq. (\ref{s4new}) below:


\begin{widetext}


\begin{equation} 
\sum_{r= -\infty}^{r=\infty} \sum_{s= -\infty}^{s=\infty}
\frac{1}{x^2 (2 r + 1)^2 + s^2} \approx
 \frac{2 \pi}{x} \left[\sum_{r=0}^{\infty} \frac{\coth (( 2 r + 1)\pi x ) -1}{2 r + 1} 
+ \frac{\ln r_{max}}{2} + \frac{C}{2}\right], \label{b14}
\end{equation}

\begin{equation} 
\sum_{r= -\infty}^{r=\infty} \sum_{s= -\infty}^{s=\infty}
\frac{1}{x^2 (2 r + 1)^2 + ( 2 s+ 1)^2} \approx
 \frac{ \pi}{x} \left[\sum_{r=0}^{\infty} \frac{\tanh (( 2 r + 1)\frac{\pi
x}{2} ) - 1}{2 r + 1} + \frac{\ln r_{max}}{2} + \frac{C}{2}\right]. \label{b15}
\end{equation}


\end{widetext}

 
\noindent In Eqs.(\ref{b14}) and (\ref{b15}), we use $\sum_{s=0}^{\infty}
1/(y^2 + ( 2 s + 1 )^2) = \pi \tanh(\pi y/2)/(4 y)$. In the presence of
$\sin^2 ( (2 r + 1 ) y)$ or $\cos^2 ( (2 r + 1 ) y)$, the series can be
calculated in a similar way, using $\sin^2 ( (2 r + 1 ) y) = (1 - \cos(2 (
2 r + 1)y))/2$ or $\cos^2 ( (2 r + 1 ) y) = (1 + \cos(2 ( 2 r + 1)y))/2$.
For example,


\begin{widetext}


\begin{eqnarray} 
\sum_{r= -\infty}^{r=\infty} \sum_{s= -\infty}^{s=\infty}
\frac{\sin^2(( 2 r + 1) y)}{(x^2 (2 r+ 1)^2 + s^2)}
 &=& \frac{2 \pi}{x} \Bigg[\sum_{r=0}^{\infty} \frac{\sin^2 ( (2 r + 1 ) y
)(\coth (( 2 r + 1)\pi x ) - 1)}{2 r + 1}\nonumber \\ 
&+& \frac{\ln r_{max}}{4} - \frac{\ln |\cot(y/2)|}{4}+ \frac{C}{4} \Bigg ], 
\end{eqnarray}

\begin{eqnarray} 
\sum_{r= -\infty}^{r=\infty} \sum_{s= -\infty}^{s=\infty}
\frac{\cos^2(( 2 r + 1) y)}{(x^2 (2 r+ 1)^2 + s^2)}
 &=& \frac{2 \pi}{x} \Bigg [\sum_{r=0}^{\infty} \frac{\sin^2 ( (2 r + 1 ) y
)(\coth (( 2 r + 1)\pi x ) - 1)}{2 r + 1}\nonumber \\ 
&+& \frac{\ln
r_{max}}{4} + \frac{\ln |\cot(y/2)|}{4}+ \frac{C}{4} \Bigg ]. 
\end{eqnarray}

\noindent We use 
\begin{equation} 
\sum_{r=0}^\infty \frac{\cos((2 r +1
)\theta)}{2 r +1 } = \frac{\ln | \cot (\theta/2)|}{2}. \label{coth}
\end{equation} 



\end{widetext}




\begin{thebibliography}{10}

\bibitem{pok-rev} I.F. Lyuksyutov and V.L. Pokrovsky, Adv. Phys. {\bf54}, 67 
(2005).

\bibitem{serk-rev} S. Erdin, {\it Frontiers in Superconducting Materials}, 
edited 
by A. Narlikar, (Springer-Verlag, Berlin, 2005), pp. 425-458.

\bibitem{applphyslett} N. Touitou, P. Bernstein, J.F. Hamet, Ch. Simon {\it et al.}, 
Appl. Phys. Lett. {\bf 85}, 1742 (2004).

\bibitem{lange} M. Lange, M.J. Van Bael and V.V. Moshchalkov, Phys. Rev. B {\bf 68}, 174522 (2003).

\bibitem{dwsuper} A. Yu. Aladyshkin, A.I. Buzdin, A.A. Fraerman, A.S. Mel'nikov {\it et al.}, Phys.Rev.B {\bf 68}, 184508 (2003).

\bibitem{sonin1} E. B. Sonin, Pis'ma Zh. Tekh. Fiz. {\bf 14}, 1640 (1988) 
[Sov. Tech. Phys. Lett. {\bf 14}, 714 (1988)]

\bibitem{helseth} L.E. Helseth, P.E. Goa, H. Hauglin, M. Baziljevich, and 
T.H. Johansen, Phys. Rev. B {\bf 65}, 132514 (2002).

\bibitem{laiho} R. Laiho, E. Lahderanta, E.B. Sonin and K.B. Traito, Phys. 
Rev. B {\bf 67}, 144522 (2003). 



\bibitem{pok2} I.F. Lyuksyutov and V.L. Pokrovsky, cond-mat/9903312 
(unpublished).

\bibitem{pok7} I.F. Lyuksyutov and V.L. Pokrovsky, Modern. Phys. Lett. B
{\bf 14}, 409 (2000).


\bibitem{stripe} S.Erdin, I.F. Lyuksyutov, V.L.
Pokrovsky and V.M.
Vinokur,
Phys. Rev. Lett. {\bf 88}, 017001
(2002).


\bibitem{abrikosov}  A.A. Abrikosov, {\it Introduction to the Theory of
Metals} (North Holland, 1986).

\bibitem{serkan1} S. Erdin, A.M. Kayali, I.F. Lyuksyutov and V.L.
Pokrovsky, Phys.Rev. B {\bf 66}, 014414 (2002).


\bibitem{ser-physica} S. Erdin, Physica C
{\bf 391},140 (2003).


\bibitem{karap} G. Karapetrov, J. Fedor, M. Iavarone, D. Rosenmann and W.K. Kwok,
Phys. Rev. Lett. {\bf 95}, 167002 (2005).

\bibitem{chains} S.J. Bending and J.W. Dodgson, J.Phys: Condens. Matter. 
{\bf 17}, R955 (2005). 




\bibitem{abrom} M. Abramowitz and I. A. Stegun, {\it Handbook of
Mathematical
Functions}, (Dover Publications, New York, 1970).





\bibitem{arfken} G.B. Arfken and H.J. Weber, {\it Mathematical Methods for
Physicists},  (Academic Press,  Orlando, 2000), 5th ed.

\bibitem{series} E.R. Hansen, {\it A Table of Series and Products},
(Prentice-Hall, Englewood Cliffs, 1975).

\end{thebibliography}
\end{document}